# Kinematical coincidence method in transfer reactions

L.Acosta[2], F.Amorini[2], L.Auditore[4], I.Berceanu[8], G.Cardella[1,*], M.B.Chatterjiee[9], E.De Filippo[1], L.Francalanza[2,3], R.Gianì[2,3], L.Grassi[1,11], A. Grzeszczuk[10], E.La Guidara[1,7], G.Lanzalone[2,5], I.Lombardo[2,6], D.Loria[4], T.Minniti[4], E.V.Pagano[2,3], M.Papa[1], S.Pirrone[1], G.Politi[1,3], A.Pop[8], F.Porto[2,3], F.Rizzo[2,3], E.Rosato[6], P.Russotto[2,3], S.Santoro[4], A.Trifirò[4], M. Trimarchi[4], G.Verde[1], M.Vigilante[6]

1 INFN - Sezione di Catania, Via S. Sofia, 95123 Catania, Italy

2 INFN - Laboratori Nazionali del Sud, Via S. Sofia, Catania, Italy

3 Dip. di Fisica e Astronomia, Università di Catania, Via S. Sofia, Catania, Italy

4 INFN Gruppo collegato di Messina and Dip. di Fisica, Università di Messina, Italy

5 Facoltà di Ingegneria e Architettura, Università Kore, Enna, Italy

6 Dipartimento di scienze Fisiche, Università Federico II and INFN Sezione di Napoli, Italy

7 Centro Siciliano di Fisica Nucleare e Struttura della Materia, Catania, Italy

8 Institute for Physics and Nuclear Engineering, Bucharest, Romania

9 Saha Institute for Nuclear Physics, Kolkata, India

10 Institut of Physics, University of Silesia, Katowice, Poland

11 Rudjer Boskovic Inst., Zagreb, (Croatia)

*corresponding author

**Abstract**

A new method to extract high resolution angular distributions from kinematical coincidence measurements in binary reactions is presented. Kinematic is used to extract the center of mass angular distribution from the measured energy spectrum of light particles. Results obtained in the case of $^{10}Be+p\rightarrow{}^{9}Be+d$ reaction measured with the CHIMERA detector are shown. An angular resolution of few degrees in the center of mass is obtained.

## 1. INTRODUCTION

It is well known how the measurement of angular distribution in elastic scattering and transfer reactions induced by light ions is a very useful method to extract spectroscopic information [1,2]. In the last years, these measurements have been carried out in inverse kinematics reactions induced by radioactive beams impinging on light targets by using very performing detection systems, see for instance [3-9]. One of the most important issue that one has to handle in this type of measurements is the low beam intensity. A possible solution is the one to increase the solid angle coverage, by mounting detectors very close to the target. However, depending on the adopted detector configuration, in this way the angular resolution could be poor, strongly affecting the quality of the experimental results. In this paper, we show that, by using the kinematical coincidence method [10], angular resolution of the order of 1° in the center of mass (CM) can be easily achieved without contradiction with the large coverage of the solid angle. This method allows to perform nuclear structure studies also with very powerful $4\pi$ detector systems, that are very efficient to measure kinematical correlations but are generally characterized by poor angular resolution. Also more simple silicon arrays often used as ancillary detectors for gamma ray arrays could be very efficiently used. In this paper we test the power of this method in the study of the angular distributions of the $^{10}Be+p\rightarrow{}^{9}Be+d$ reaction at 58 A·MeV. $^{10}Be$ beam was produced by using the fragmentation method at INFN Laboratori Nazionali del Sud (INFN-LNS) in Catania [11]. Reaction products were detected with the CHIMERA detector [12-13]. CHIMERA is a $4\pi$ detector with a granularity (1192 telescopes) suitable for the study of multi-fragmentation reactions between heavy ion from 10 to 100 A·MeV. The large segmentation of the apparatus allows to get, at very forward angles, a resolution better than 1°, while, at angles larger than 30°, the angular resolution is $\Delta\theta = \pm4°$. This means, for examples, that in the reaction here studied we can get approximately $\Delta\theta = \pm 8°$ resolution in the CM system for the deuterons angular distribution. The angular resolution is moreover influenced by the emittance of the beam that is induced by the angular straggling of the projectile fragmentation process in the thick production target used. We have a beam spot on the nuclear target of the order of 2x1 cm$^2$ with an angular spread of the beam around ±1° with a consequent degradation of the angular resolution. We show in the following how the used method allows also to automatically correct for the beam angular spread obtaining an angular distribution with a resolution substantially influenced only by the statistics of the measurement.

## 2. THE FRAGMENTATION BEAM CHARACTERISTICS

The fragmentation beam was produced by using an $^{18}O^{7+}$ primary beam delivered by the INFN-LNS superconducting cyclotron at 55 AMeV. The fragmentation reaction was induced on a $^{9}Be$ target 1.5 mm thick mounted in the first section of the transport beam line. Following LISE++ simulations [14], the beam line was set to maximize production of $^{11}Be$ ions ($B\rho$=2.78Tm). The transport of the fragmentation beam was optimized by using the radioactive beam diagnostic system of the INFN-LNS [15]. The beam was identified in particle by particle mode by using a tagging system consisting on two double side silicon strip detectors (DSSSD) and a large surface micro-channel plate (MCP) detector [16]. The first DSSSD detector, hereafter named *tagging strip*, was 64$\mu$m thick with 24 strips on each side and a total surface of 24x24mm$^2$. It was placed 2m before the CHIMERA target and was used to measure the beam energy loss $\Delta E$, and its X-Y position. A second DSSSD detector, named *trajectory strip*, 72$\mu$m thick, 5x5cm$^2$ surface with 16 strips on each side, was placed about 80 cm before the CHIMERA target. Being very near (20cm) to the entrance hole ( 6 cm diameter) of the CHIMERA apparatus, the particles produced by reactions in such silicon detector could be a large source of spurious events. Therefore it was used only during beam transport for adjustment purposes and every 12 hours for stability check. In fig.1a) we plot a sketch (not in scale) of the strip detectors and target ( 50$\mu$m thick polyethylene $C_2H_4$) in order to define the trajectory measurement. In fig.1.b) the calculated beam profile on the target is shown, we note a small vertical misalignment and the approximate size of 2 cm along the vertical axis and of 1 cm along the horizontal one. In fig. 1.c) and d) we plot respectively the impinging angle upon the nuclear target $\theta_{beam}$ as a function of the vertical and horizontal position in the tagging strip. We note a strong correlation between $\theta_{beam}$ and $x_{strip}$, with a rather narrow distribution of this angle. This strict correlation is mainly due to the last magnetic dipole of the transport beam line. The MCP detector, 4x6cm$^2$ wide [16], was placed approximately 13 m before the tagging strip and was used as start of the time of flight (TOF) measurement of the beam particles ( the stop being delivered by the *tagging strip* ). In fig 2 we show the quality of the identification obtained by plotting the TOF as a function of the energy loss measured in the *tagging strip*. The isotopic beam identification was obtained for comparison with LISE++ predictions and it was further checked in charge identification and, when possible, in mass by looking to elastic reaction products detected by the forward telescopes of the CHIMERA array.

## 3. CALIBRATION PROCEDURES

The individual detection cell of the CHIMERA detector is a telescope constituted by a first stage silicon detector (300 μm thick). The second stage is a CsI(Tl) scintillator with photodiode readout. Different techniques are used for particle identification allowing for mass and/or charge identification in a large angular and energy range [17-21]. The energy of the impinging particle can be obtained by summing the energy deposited in both detector stages of the telescope. The silicon stage was calibrated in energy by using peaks from low energy light beams elastically scattered by thin Au targets. The situation is more complex for CsI(Tl) detectors due to quenching and non-linearity effects at low kinetic energy in the energy response function. A first order energy calibration of CsI(Tl) in the forward rings was obtained using the elastic peaks on carbon targets measured with fragmentation beams. Also elastic peaks generated by the scattering of low energy proton beams on carbon and gold targets were used. In fig. 3 such calibration points for one detector are plotted as energy against the CsI(Tl) total light collected (*Fast* variable in channels) [18,19]. In order to take into account the charge and mass dependence of the energy-light response function of detectors we have used the formula suggested by Horn [22]

$$L=a_1\{E-a_2AZ^2\ln[(E+a_2AZ^2)/(a_2AZ^2)]\}+a_0$$

where A, Z and E are respectively the mass, charge and energy of the detected fragment, L is the collected light signal, $a_0$ is a parameter depending on pedestal, $a_1$ is connected to the electronic gain of the channel and include also the scintillation efficiency of the detector, $a_2$ is related to the Birks quenching factor [23]. This formula is based on the assumption that the quenching of CsI(Tl) light output depends on the specific energy loss of the particle dE/dx (Birks prescription). As can be observed in fig. 3, the fitting parameters allow the reproduction of the general behavior of the experimental response function. After this first order calibration, only second order corrections (few percent) were included, when necessary, for each charge and detector, to better reproduce the energy of elastic and inelastic peaks observed. These corrections take into account for small differences in the crystal doping and wrapping, photodiode coupling, and electronics response.

At more backward angles, where practically only light particles were detected, low energy proton elastically scattered on various targets were used for energy calibration. The parameters extracted from the fit of the most forward telescopes and previous data with deuteron beams [23] confirm us that this calibration can be extended under suitable approximations to all hydrogen isotopes[24].

At very forward angles the standard CHIMERA charge preamplifiers have a conversion sensitivity of 2mV/MeV, in order to avoid saturation effects due to the expected large dynamical range. Evidently, for the lightest particles this fact does not allow for a clear isotopic identification by using the ΔE-E method, as instead obtained at larger angles [17], so apart few very well performing telescopes, in the forward direction we had only charge identification for the heavy reaction partners as shown in fig.4.

Particles emitted at angles larger than 20° were identified in charge and mass with $\Delta$E-E method and for Z≤2 also with fast-slow method, fig.5

## 3. DATA ANALYSIS AND RESULTS

As seen in Fig. 2 a "cocktail" of fragmentation beams are identified in the tagging strip. Thus the first step of the data analysis consists on the selection of the beam under study ($^{10}$Be in this case ) with the use of appropriate cuts. Events were further selected, searching for d-Be coincidences with Beryllium ions (charge identification) in the most forward rings and deuteron ions ( isotopic identification ) in the angular range which was allowed by kinematics. Only events with charged particles multiplicity equal to two hits were analyzed, strongly reducing contaminations due to carbon in the polyethilene target, and to reactions in the *tagging detector*. Other constraints were taken into account by using conservation laws. Firstly, due to momentum conservation the relative azimuthal angle $\Delta\phi$ between the two fragments must be 180°. In fig. 6, we plot this angle as measured for the investigated reaction, $^{10}$Be+p→$^{9}$Be+d. We can recognize the binary events concentrated in the peak around 180°. The width of this peak is about +-20°, due to the $\phi$ opening angle of detectors.

A further selection, based on energy conservation law, requires that the sum of kinetic energies of the two detected particles is equal to the beam energy plus the Q-value (Q=-4.58 MeV), see fig.7. Notice in this plot a peak close to the value of 580 MeV, that roughly corresponds to the total available energy. Due to the relatively poor CsI(Tl) energy resolution for heavy fragments, it is quite difficult to discriminate the decay path towards to the ground state of a specific nucleus with respect to excited states by just looking only at the peak in the total energy spectrum. However, in the simple case here investigated, the neutron separation energy for $^{9}$Be is only 1.665 MeV, and even the first excited level (1.684 MeV) is unbound, decaying to the n+2$\alpha$ channel. Therefore, by requiring a beryllium in the final channel we rejected excited level in a natural way. We conclude that we are observing only the GS level.

In fig.8 we can finally plot the deuteron energy spectrum as it was measured for this channel. The low energy part of the spectrum is affected by efficiency problems better described below. It is interesting to note the presence of two relative minima near 20 and 50 MeV. Because in this case the final channel is well defined, we can easily convert this energy spectrum in the CM angular distribution by using the kinematic relation, computed using LISE++, shown in fig.9. By using such relation we can convert the $\Delta$E energy interval of each channel in fig.8 in angular interval $\Delta\theta_i$ directly in the CM system. With this method we can therefore deduce the number of particles detected in the $\Delta\theta_i$ angular range. Dividing this number for the solid angle subtended by the arc of sphere $\Delta\theta_i$ and taking into account the number of the beam particles ($7.3\cdot10^8$ ) and the areal density of the target nuclei, we can get the absolute cross section.

We underline that using the deuteron energy to determine the center of mass angle, we automatically correct for the spread of the beam impinging angle described in paragraph 2, fig.1.

Obviously we have to include also efficiency corrections for poorly performing detectors computed by looking at the rings where coincidence events were detected. We have to further note that in this experiment some CHIMERA rings between 7° to 20° degrees in the laboratory frame were missing because in use to another experiment [26]. Thus a more relevant correction has to be introduced. According to the simulations, 100% efficiency was maintained only from $\theta_{cm}\approx20°$ to $\theta_{cm}\approx60°$. However due to the fragmentation beam angular spread, already included in the simulation, the efficiency decreasing with the angle is rather smooth and we can observe coincidences in the CM angular range from approximately 15° to 70°. In fig.10 we plot the efficiency corrected angular distribution (full dots). The minima evidenced in the deuteron energy spectrum are clearly converted into minima in the angular distribution. The value of the size of the angular bins of each point was mainly governed by the need to get a reasonable statistical error. To our knowledge, there are no previous data on this angular distribution available in the literature. We can however compare it with the angular distribution measured by Auton [27], recently reanalyzed in [28], for the reaction $d+{}^{10}Be \rightarrow t+{}^{9}Be_{gs}$ at 15 MeV deuteron beam energy. These data are also reported in fig.10 as full squares. A cross section more than 3 times larger was measured with deuteron beam. This is consistent with the larger CM energy available in our experiment. The minima present in our data are similar to those observed by Auton even if at slightly shifted angles. Data in ref. [auton] were reproduced assuming L=1 angular momentum transfer. Due to the spins and parities involved in the studied reaction the same transferred angular momentum is expected, therefore the observation of a similar behavior in the angular distribution is understandable. DWBA calculations are in progress and will be presented in future work.

## CONCLUSIONS

We have shown that detailed angular distributions can be extracted in binary reactions induced by exotic nuclei impinging on light targets by using the kinematical coincidence method. The deuteron energy resolution of our data is of the order of 1 MeV as evaluated by elastic scattering of protons and the approximation on the deuteron energy calibration is of the same order of magnitude. This energy resolution in the light particle energy spectrum is enough to induce a CM angular resolution better than 1° due to the approximately linear correlation over a large angular range between such two quantities. This method has the great advantage to automatically correct for the angular spread of the impinging fragmentation radioactive beam. Note also that the relatively large energy spread of the fragmentation beam ($\Delta P/P$= 1%) produces in our case a very small effect, seen only at very backward CM angles, see fig.9 dashed line. This effect was neglected for the purpose of this paper. This method does not need a high resolution in the measurement of the energy of the heavy fragment, around Fermi energy, if nuclei with only unbound excited levels are investigated as in the case of the ${}^{9}Be$. The method can be however extended also to the case of bound excited levels if coincidence gamma ray measurements are performed allowing for discrimination of the different contribution of the decay process.

Thanks are due to Dr.A.Pagano for various discussions and his help and encouraging suggestions since the beginning of this work.

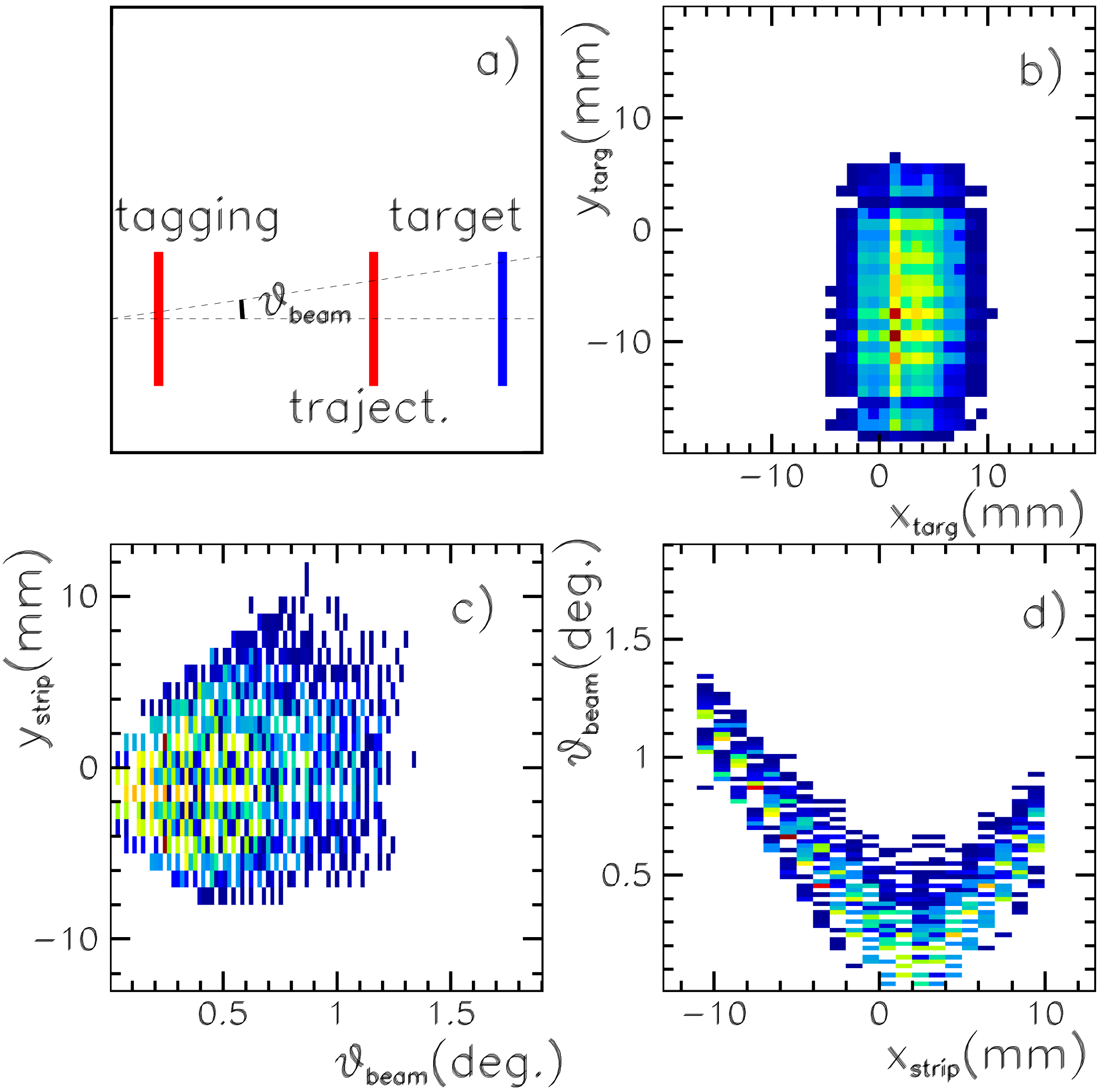

**Fig. 1a) Sketch not in scale of the trajectory measurement; b) beam image on target; c) beam impinging angle as a function of its vertical position on the tagging strip; d) Beam impinging angle as a function of its horizontal position in the tagging strip. A clear correlation is observed**

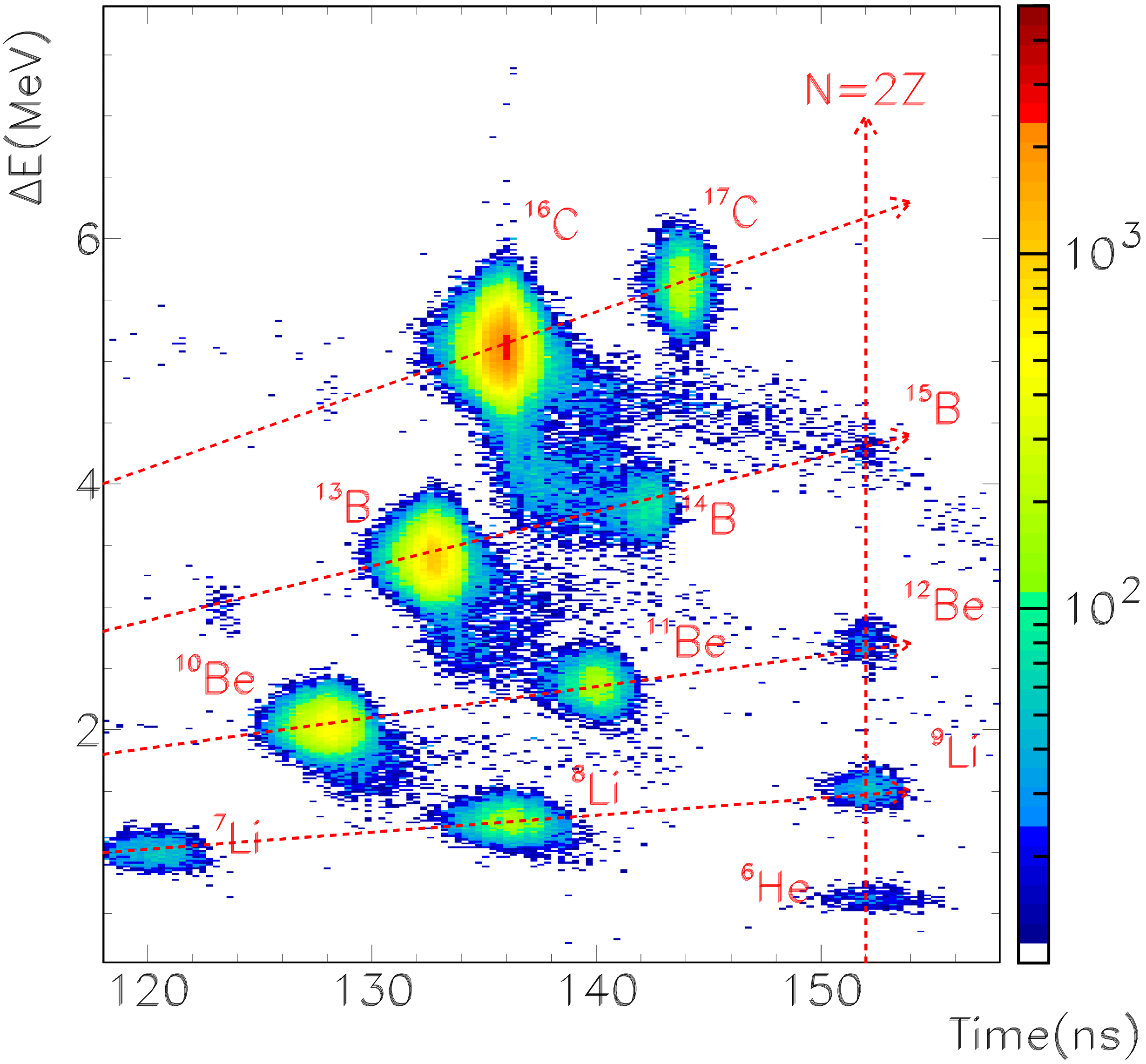

**Fig. 2ΔE-TOF identification scatter plot of the fragmentation beam used. The arrows show the loci of the different isotopes.**

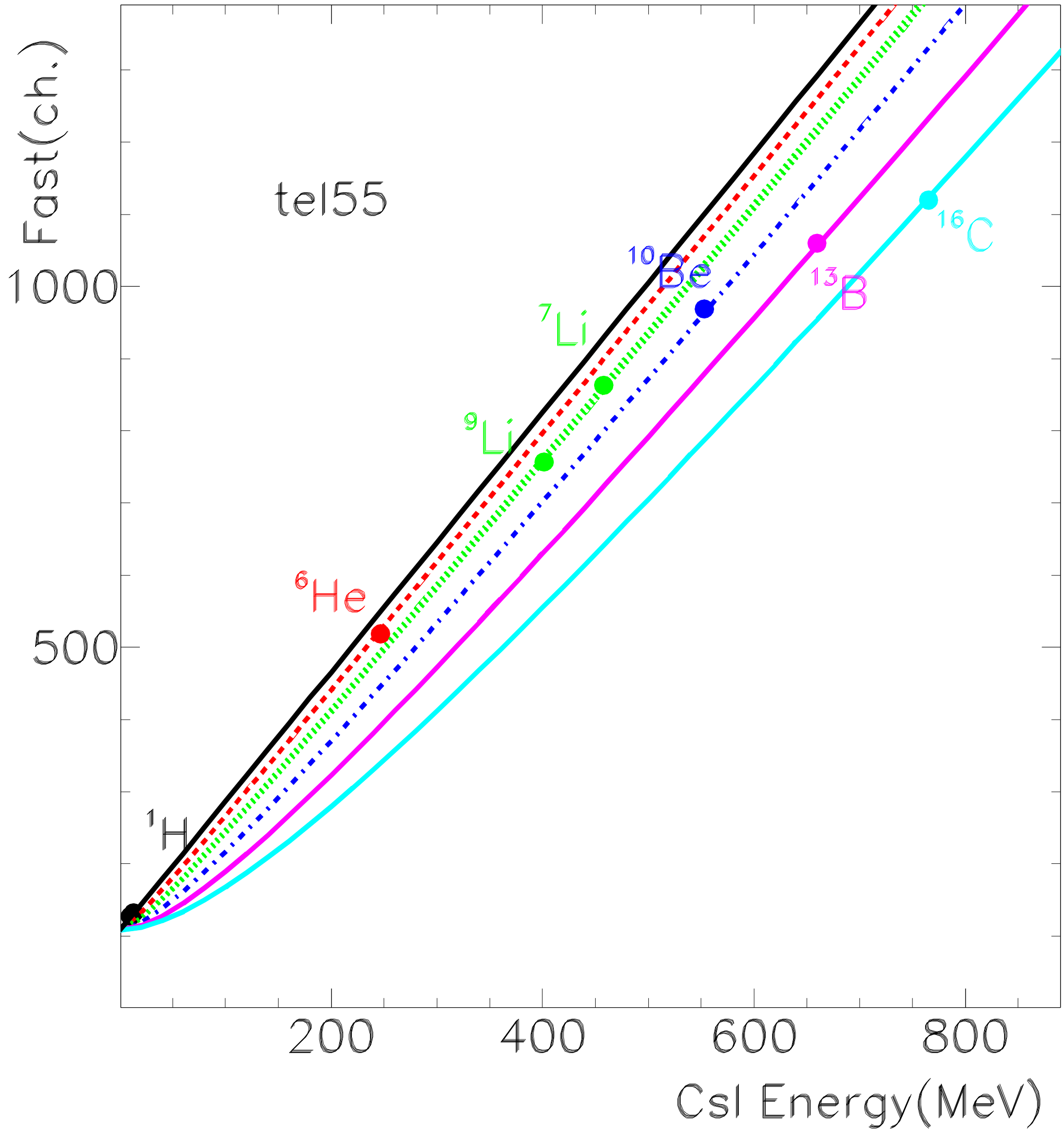

**Fig. 3 Calibration points and fit results for one CsI(Tl) detector of the CHIMERA forward rings**

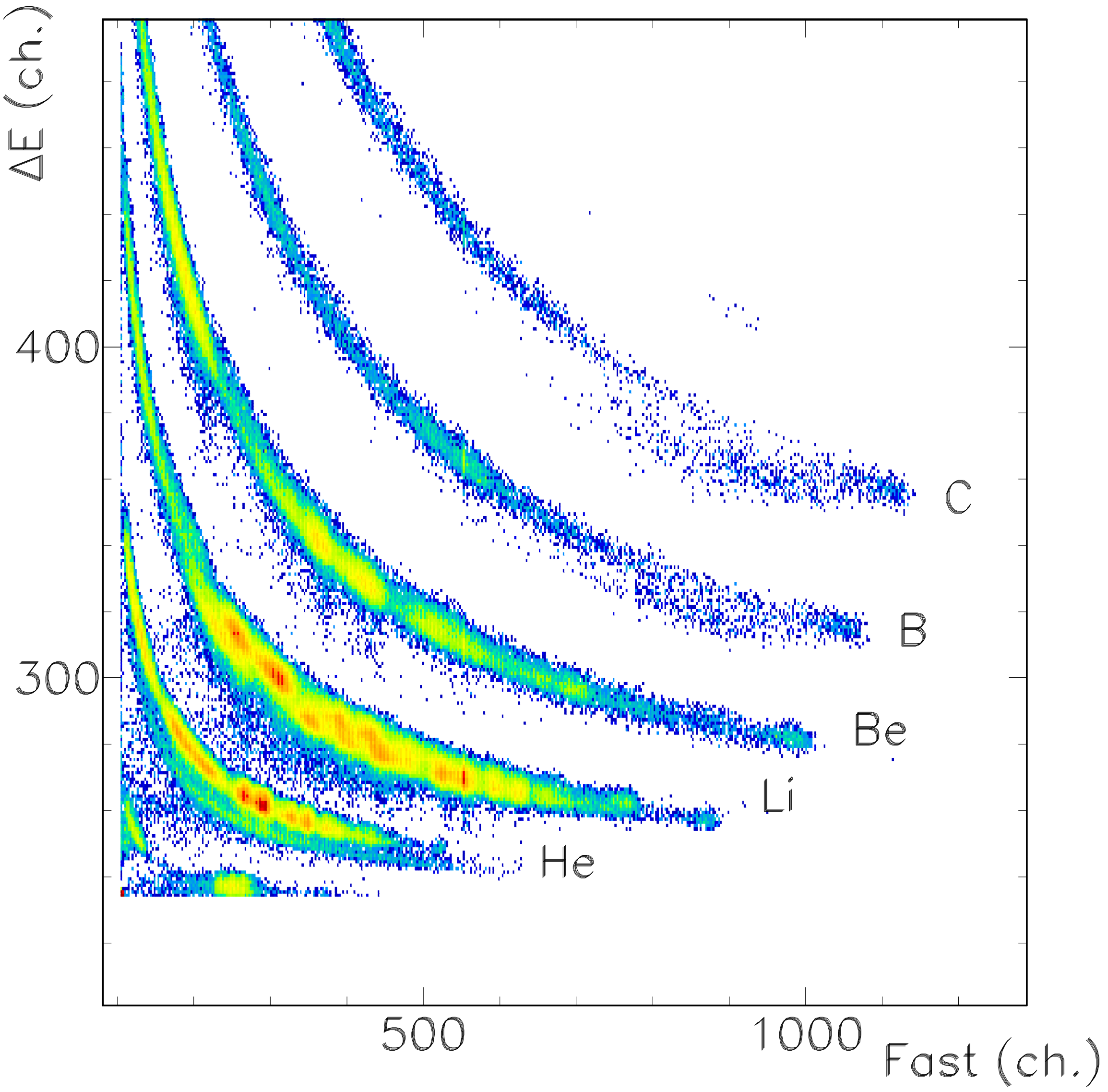

**Fig. 4 ΔE-E scatter plot of a telescope at 4.1° measured with all fragmentation beams on plastic target**

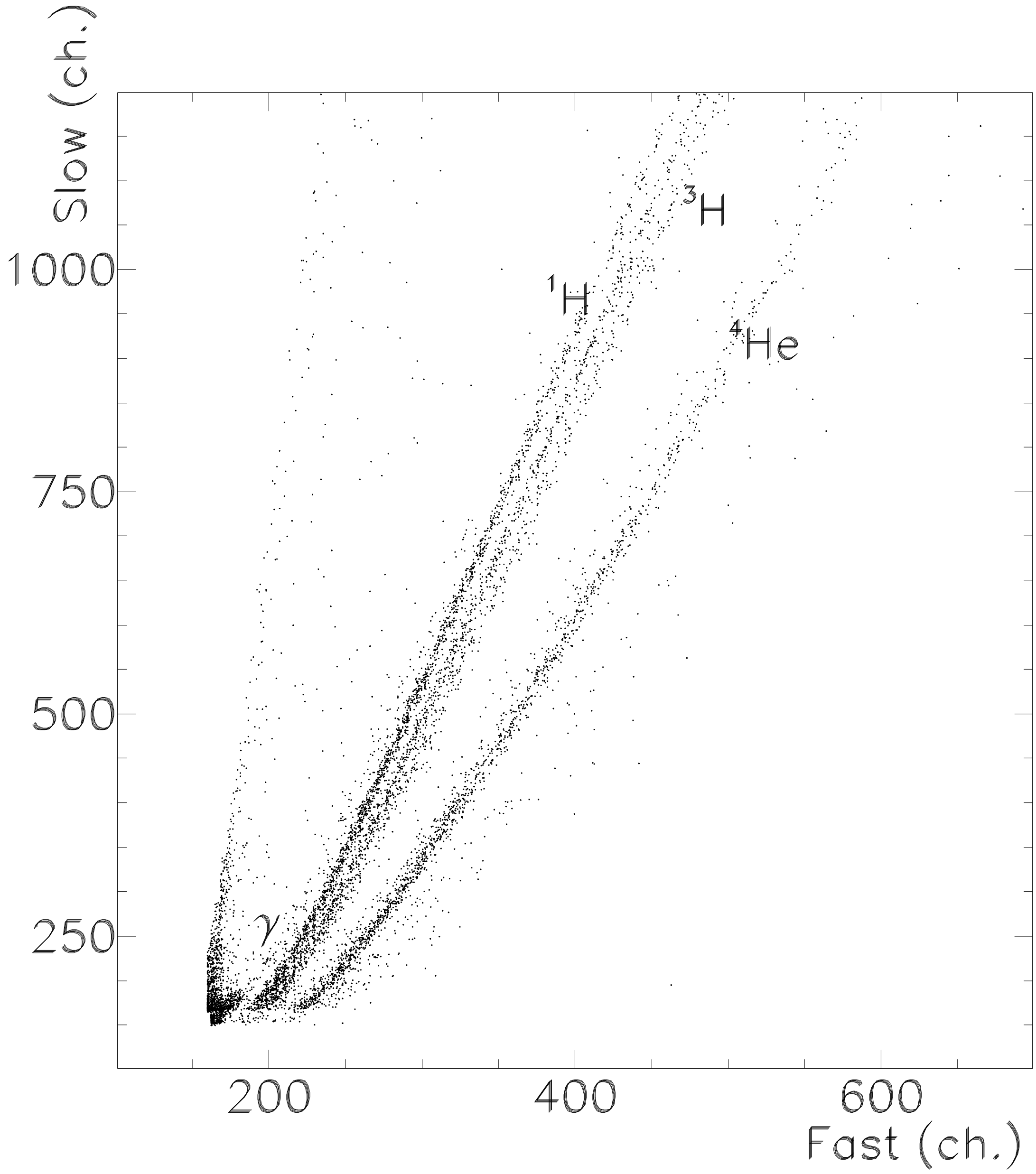

**Fig. 5 Fast Slow scatter plot of a telescope at 34°**

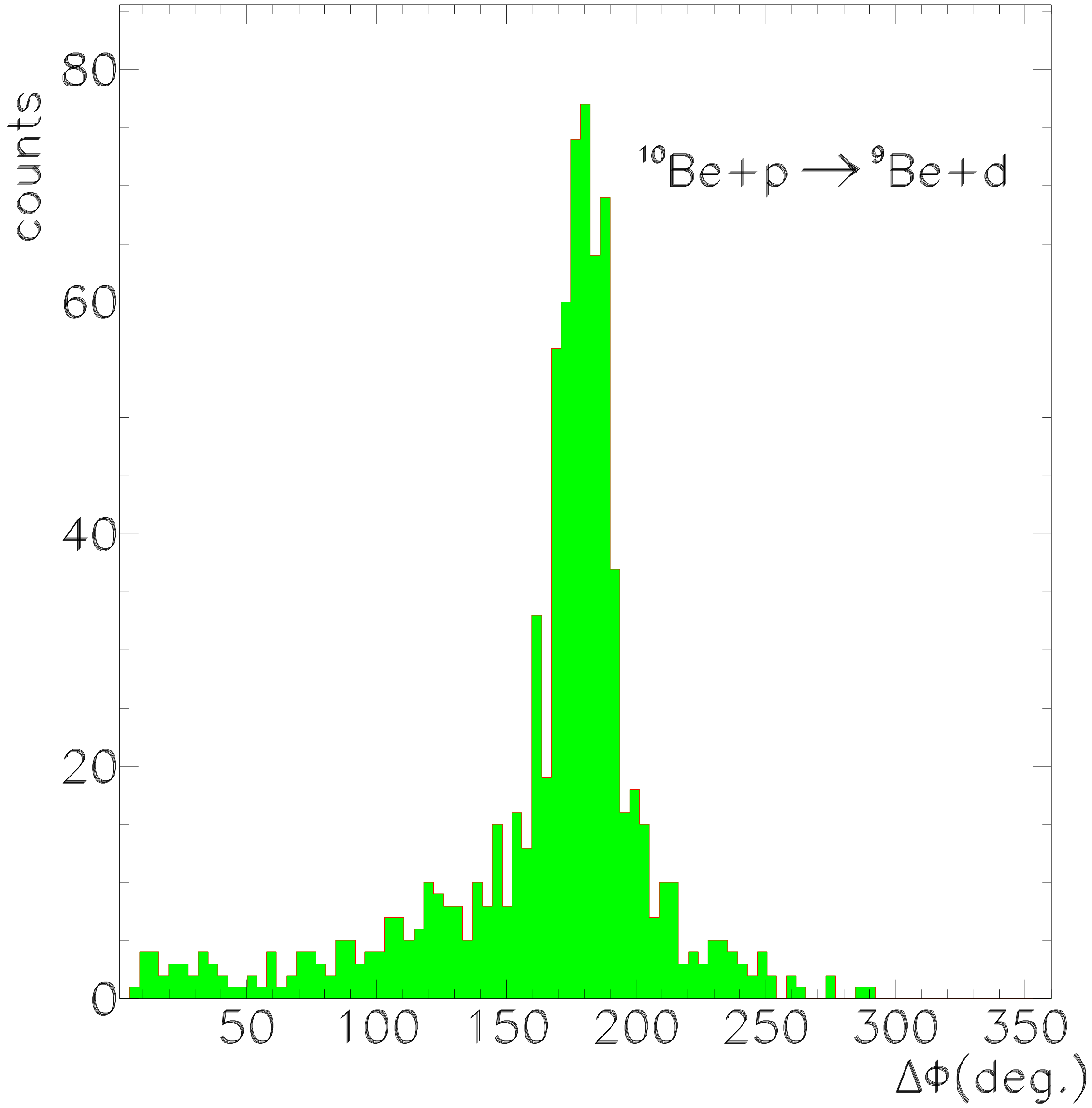

**Fig. 6 Relative angle Δϕ between the telescopes selected in coincidence. The peak at 180° is due to kinematical coincidences.**

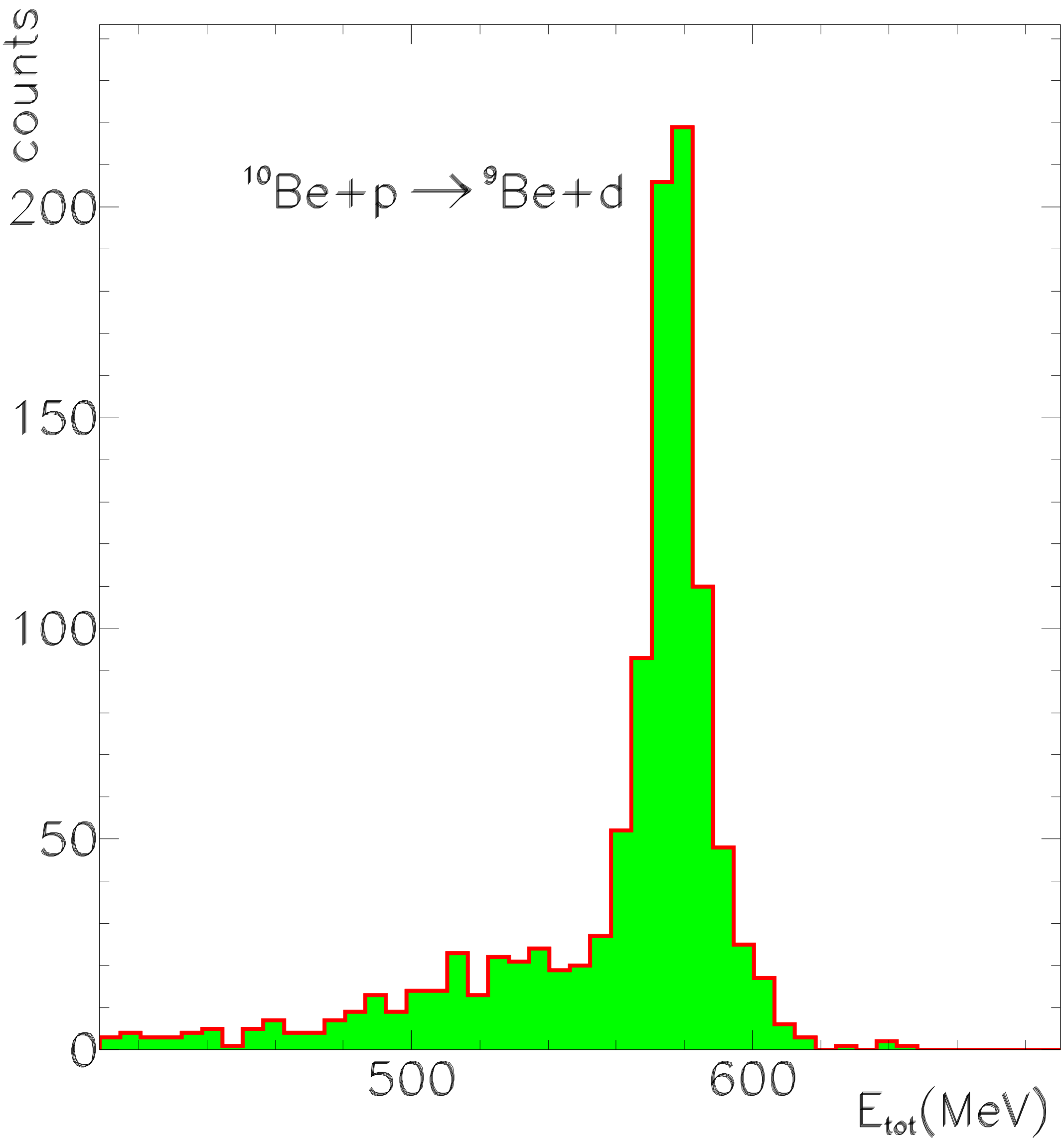

**Fig. 7 Total kinetic energy detected in the reaction 10Be+p→9Be+d**

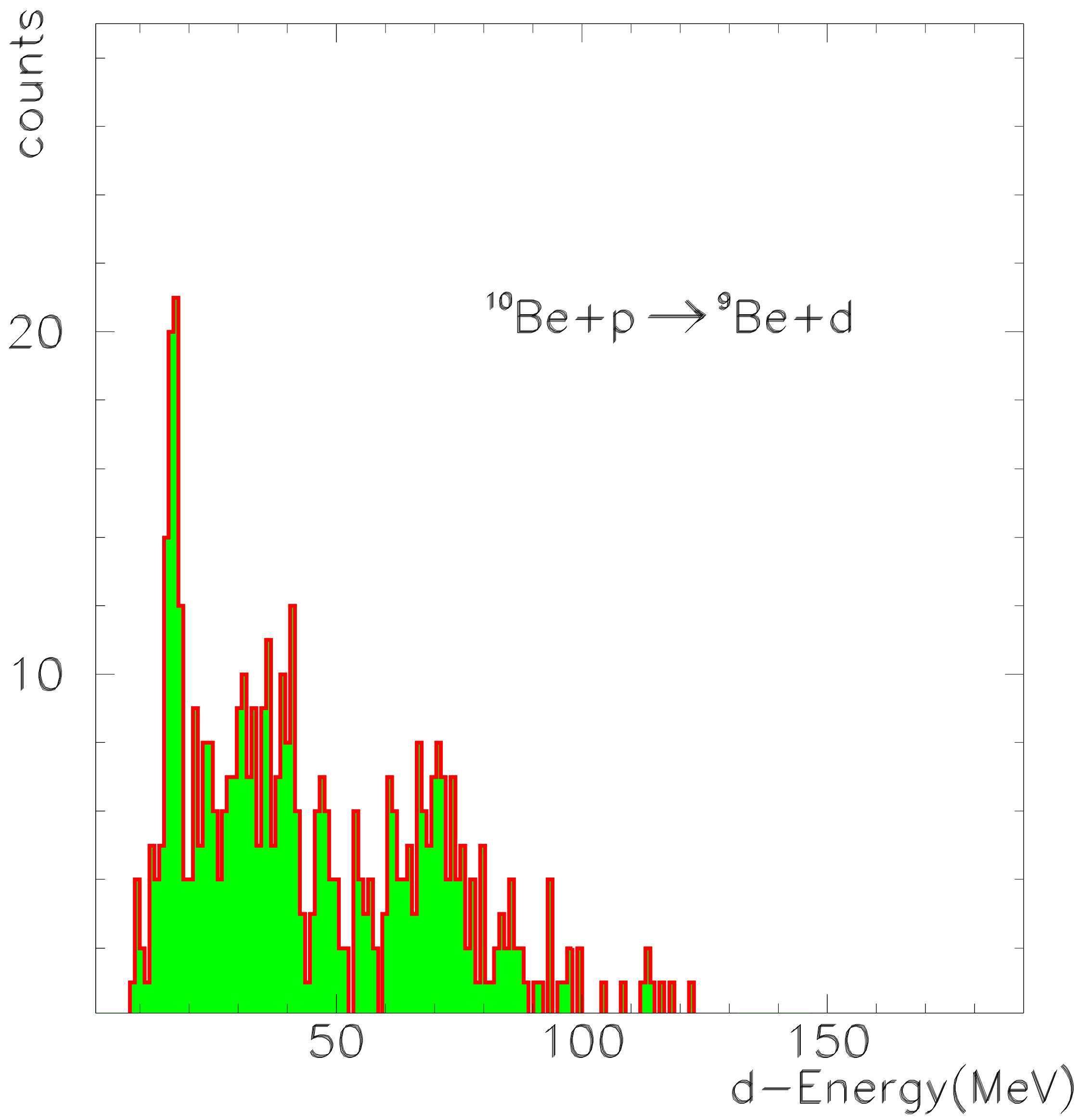

**Fig. 8 Fig.8 Deuteron energy spectrum from the reaction 10Be+p→9Begs+d**

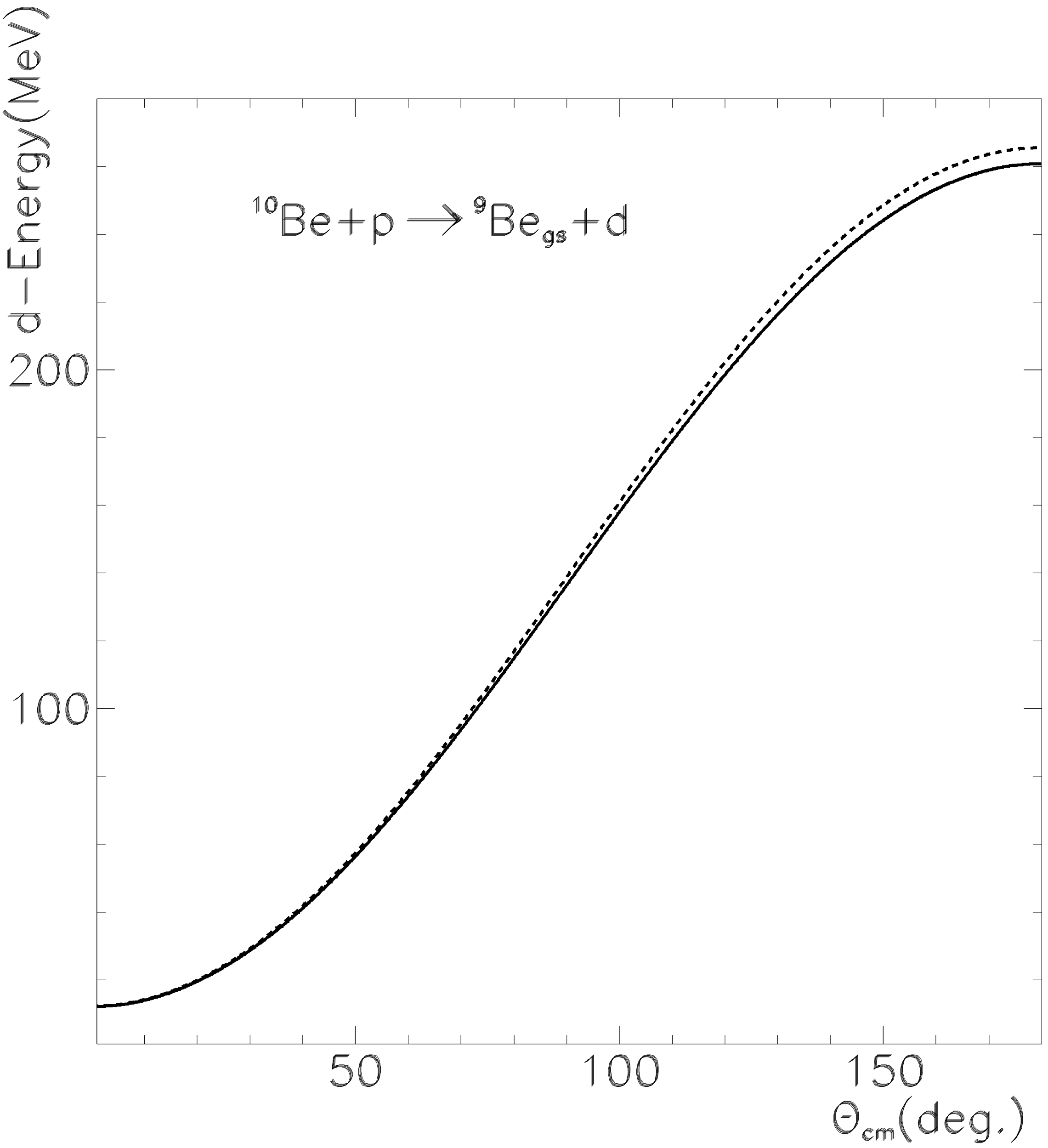

**Fig. 9 Kinematical correlation between the deuteron energy and the θcm in the reaction 10Be+p→9Begs+d 58AMeV (full line). Dashed line is computed for a beam energy of 59AMeV.**

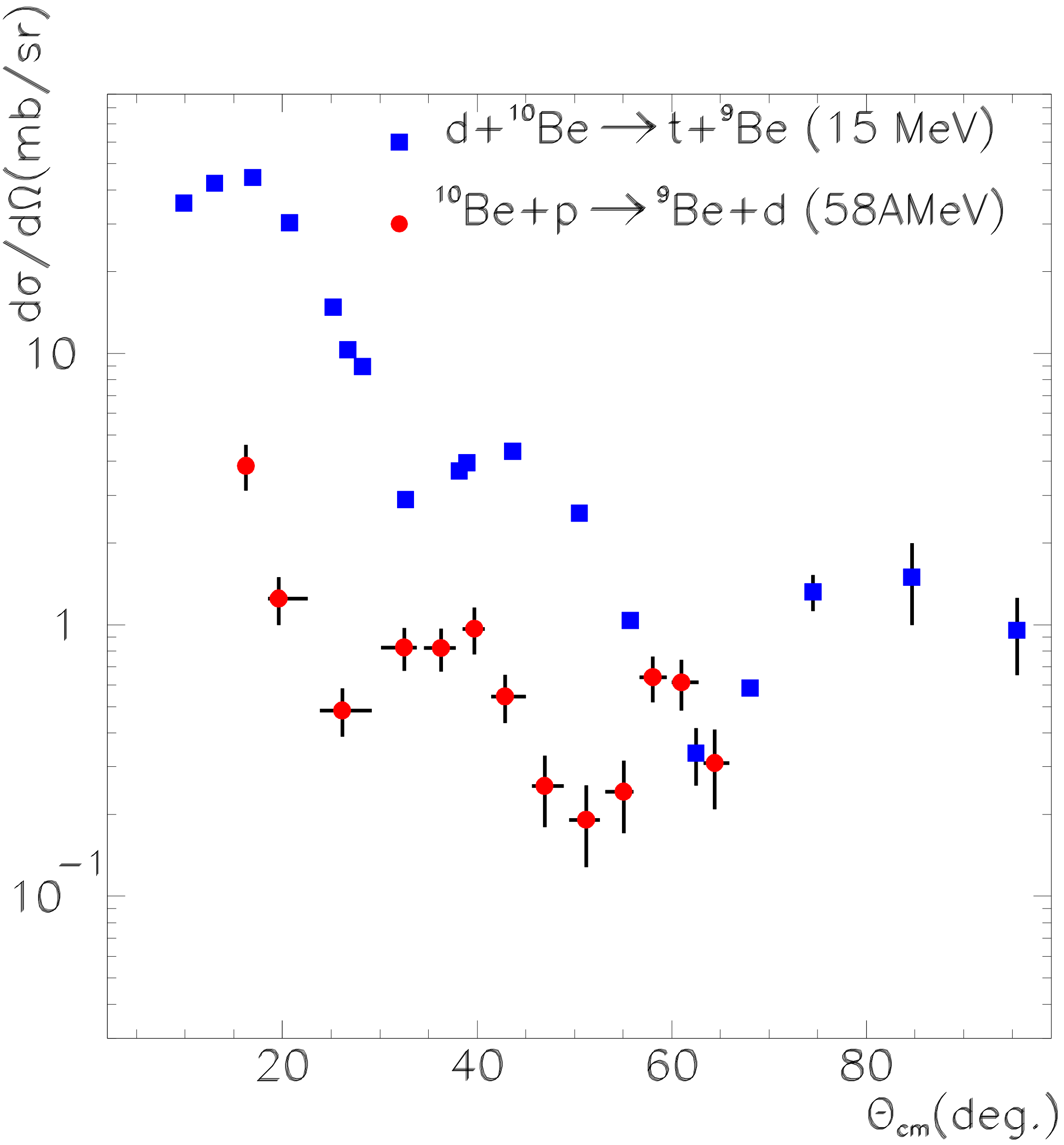

**Fig. 10 Angular distribution converted from the deuteron energy spectrum of fig.8 (full dots). Square symbols are from ref.[27].**